\documentclass[useAMS,usenatbib]{mn2e}

\usepackage{graphicx}
\usepackage{epstopdf}
\usepackage{subfigure}
\usepackage{amssymb}
\usepackage{amsbsy}
\usepackage{amsmath}    
\usepackage{listings}
\usepackage{color}
\usepackage{textcomp}
\definecolor{listinggray}{gray}{0.9}
\definecolor{lbcolor}{rgb}{0.9,0.9,0.9}
\usepackage{float}
\usepackage{comment}

\title[Spheroidal post-mergers in the local Universe]{Spheroidal post-mergers in the local Universe}
\author[A. Carpineti et al.]{Alfredo Carpineti$^{1}$\thanks{E-mail:
alfredo.carpineti07@imperial.ac.uk (AC);
sugata.kaviraj@imperial.ac.uk (SK).}, Sugata Kaviraj$^{1,2}$,
Daniel Darg$^{2}$, Chris Lintott$^{2}$, \and Kevin
Schawinski$^{3}$ and Stanislav Shabala$^{2,4}$\\ \\$^{1}$Blackett
Laboratory, Imperial College London, London SW7 2AZ,
UK\\$^{2}$Department of Physics, University of Oxford, Keble Road,
Oxford OX1 3RH UK\\$^{3}$Department of Physics,Yale University,
New Haven, CT 06511, USA\\$^{4}$School of Mathematics $\&$
Physics, Private Bag 37, University of Tasmania, Hobart 7001,
Australia}

\begin{document}

\date{Not yet published}

\pagerange{\pageref{firstpage}--\pageref{lastpage}} \pubyear{2002}

\maketitle

\label{firstpage}

\begin{abstract}
Galaxy merging is a fundamental aspect of the standard
hierarchical galaxy formation paradigm.  {\color{black}Recently,
the Galaxy Zoo project has compiled} a large, homogeneous
catalogue of 3373 mergers, through direct visual inspection of the
entire SDSS spectroscopic sample. We explore a subset of galaxies
from this catalogue that are spheroidal `post-mergers' (SPMs) -
where a single remnant is in the final stages of relaxation after
the merger and shows evidence for a dominant bulge, making them
plausible progenitors of early-type galaxies. {Our
results indicate that the SPMs} have bluer colours than the
general early-type galaxy population possibly due to merger-induced star
formation. An analysis using optical emission line ratios
indicates that {20 of our SPMs exhibit LINER or
Seyfert-like activity (68$\%$), while the remaining 10 galaxies
are classified as either star forming (16$\%$) or quiescent
(16$\%$)}. A comparison to the emission line activity in the
ongoing mergers from Darg et al. indicates that the AGN fraction
rises in the post-mergers, suggesting that \emph{the AGN phase
probably becomes dominant only in the very final stages of the
merger process.} The optical colours of the SPMs and the plausible
mass ratios for their progenitors indicate that, while a minority
are consistent with major mergers between two early-type galaxies,
the vast majority are remnants of major mergers where at least one
progenitor is a late-type galaxy.
\end{abstract}


\section{Introduction}
\begin{figure*}
\centering
\begin{tabular}{c c c c}
\includegraphics[width=1.1in]{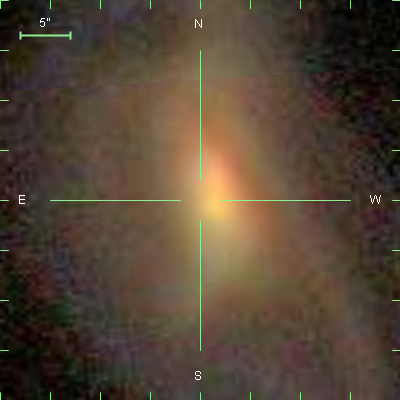} & \includegraphics[width=1.1in]{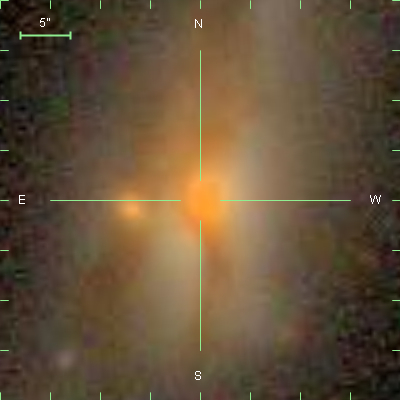} & \includegraphics[width=1.1in]{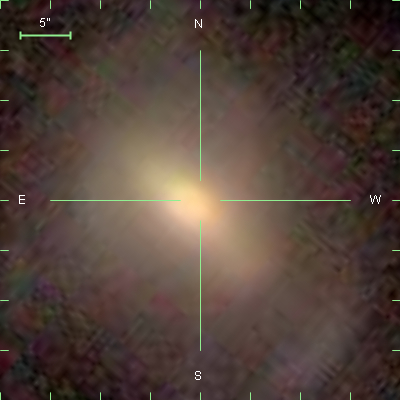} &\includegraphics[width=1.1in]{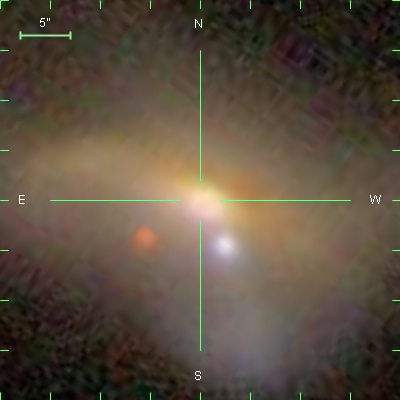}\\
\includegraphics[width=1.1in]{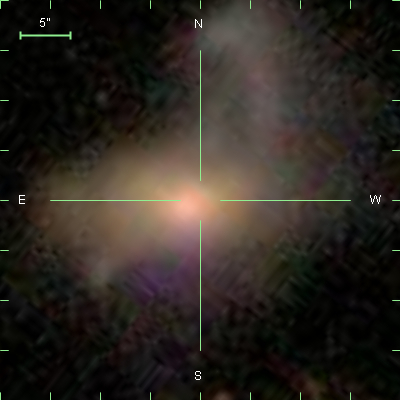} & \includegraphics[width=1.1in]{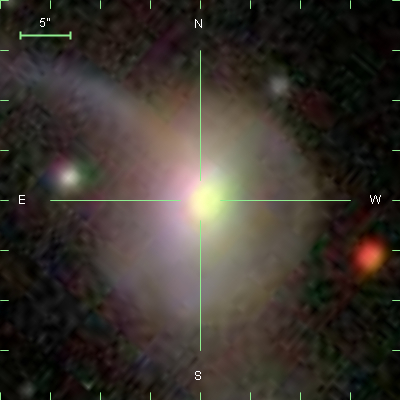} & \includegraphics[width=1.1in]{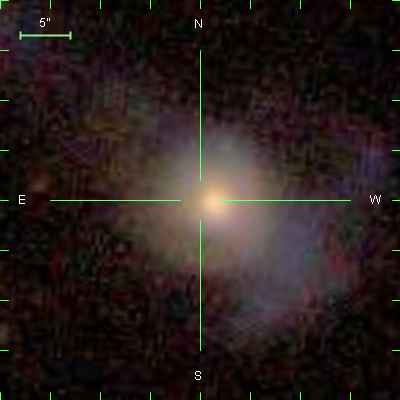} & \includegraphics[width=1.1in]{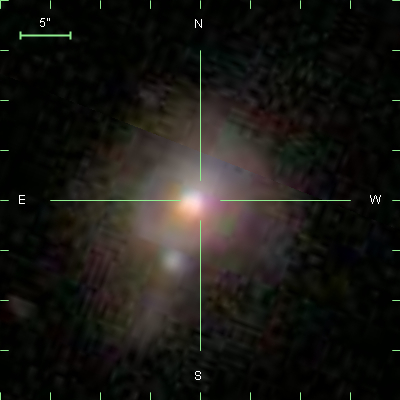}\\
\includegraphics[width=1.1in]{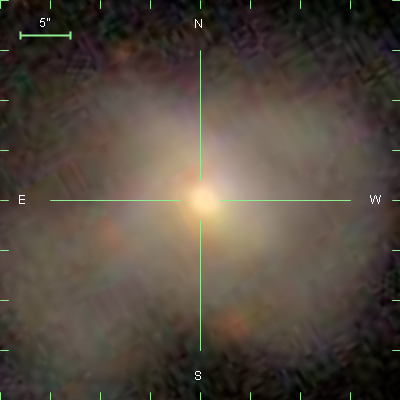} & \includegraphics[width=1.1in]{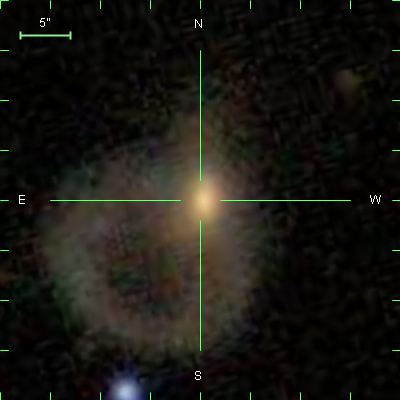} & \includegraphics[width=1.1in]{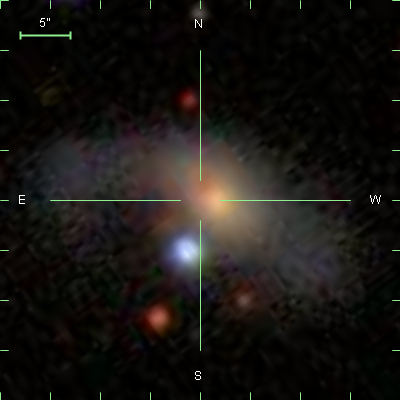} & \includegraphics[width=1.1in]{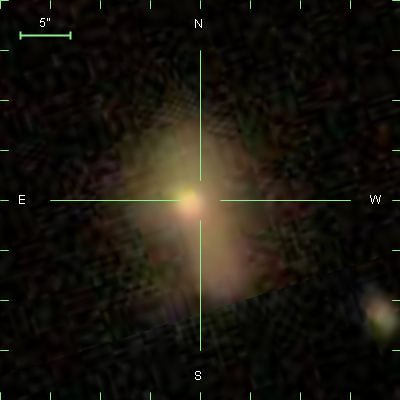}\\
\includegraphics[width=1.1in]{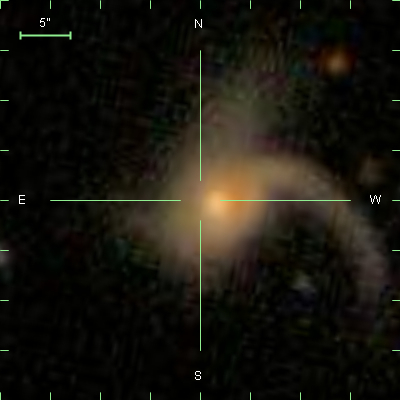} & \includegraphics[width=1.1in]{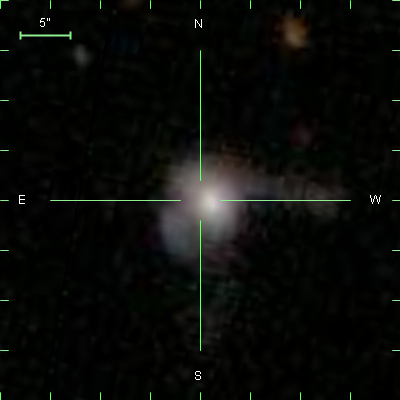} & \includegraphics[width=1.1in]{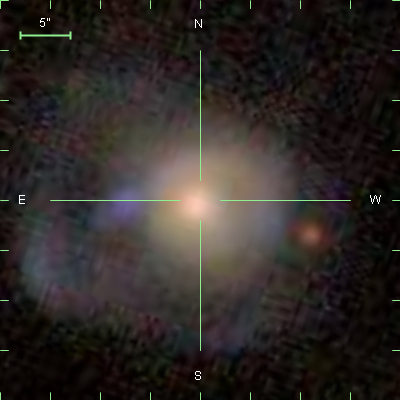} & \includegraphics[width=1.1in]{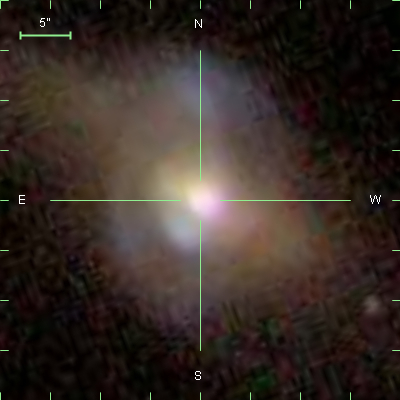}\\
\includegraphics[width=1.1in]{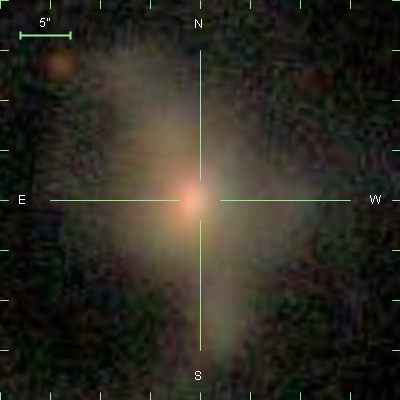} & \includegraphics[width=1.1in]{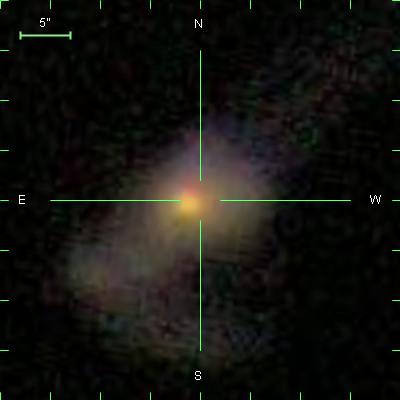} & \includegraphics[width=1.1in]{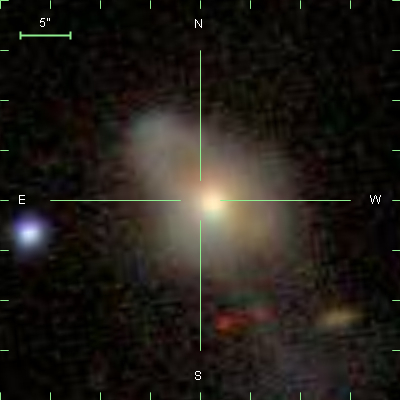} & \includegraphics[width=1.1in]{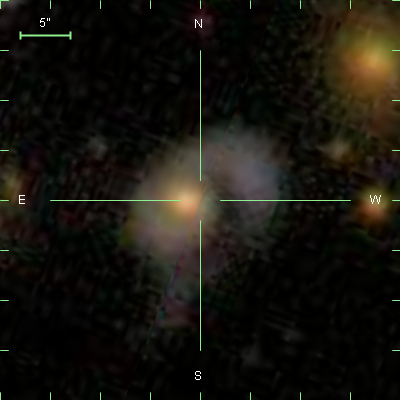}\\
\includegraphics[width=1.1in]{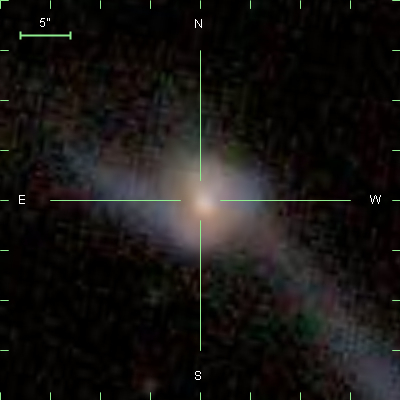} & \includegraphics[width=1.1in]{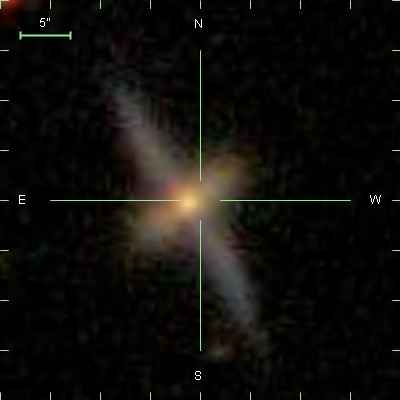} & \includegraphics[width=1.1in]{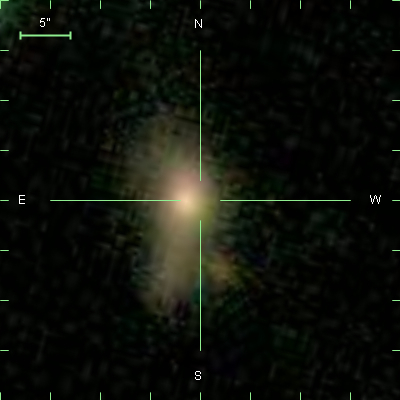} & \includegraphics[width=1.1in]{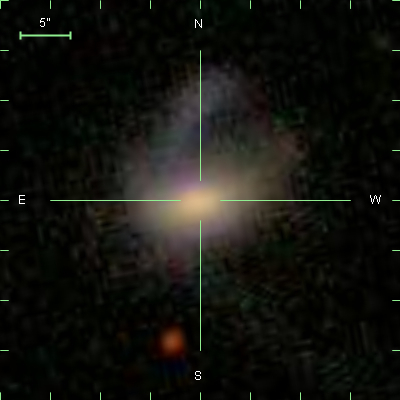} \\
\includegraphics[width=1.1in]{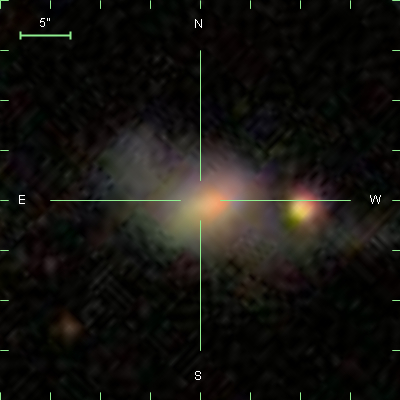} & \includegraphics[width=1.1in]{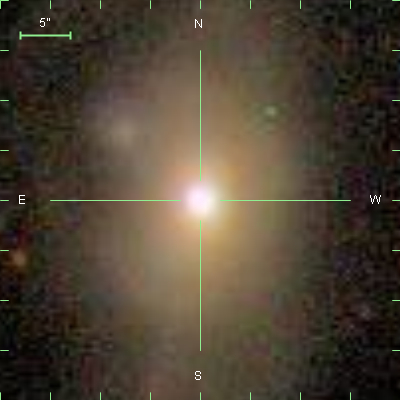} & \includegraphics[width=1.1in]{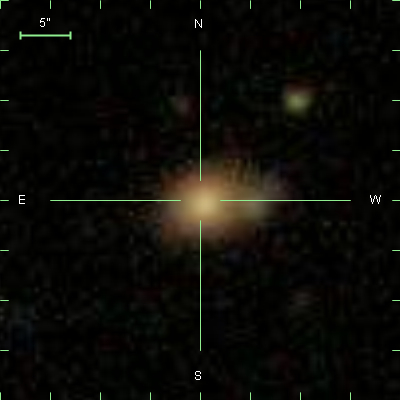} &\includegraphics[width=1.1in]{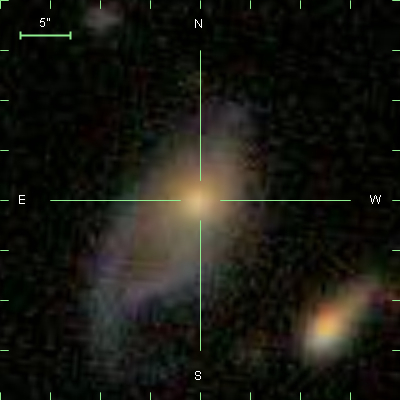}\\
\includegraphics[width=1.1in]{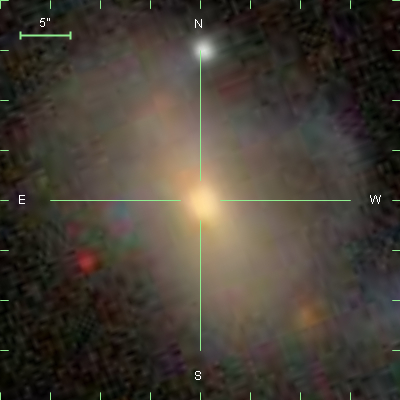} & \includegraphics[width=1.1in]{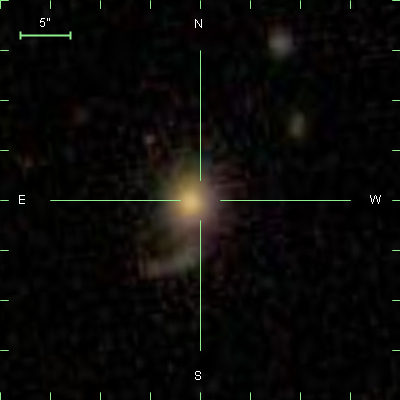} \\
\end{tabular}
\caption{\small SDSS images of our sample of spheroidal
postmergers.\label{fig:sample}}
\end{figure*}

Studying how galaxies form and evolve is a fundamental step to
better understanding our place in the Universe. The Universe is
believed to follow a $\Lambda$CDM cosmology \citep[e.g.][]{Blu84,
Fre01,Efs02, Pry02, Spe03} in which $\sim70\%$ is composed of dark
energy, while the remaining $\sim$30\% is made of matter,
subdivided into baryonic($\sim5\%$) and dark($\sim25\%$). This
model appears consistent with experimental measurements of the
Cosmic Microwave Background (CMB) \citep[e.g.][]{Dun09,Mat09} and
large scale clustering \citep[e.g.][]{San09}. A key feature of the
$\Lambda$CDM cosmogony is a hierarchical bottom-up formation
paradigm, with smaller bodies accreting to form progressively
larger ones \citep{Whi78,Sea78}. Small dark matter halos form
first and subsequently merge to form bigger halos
\citep[e.g.][]{Pee82, Blu84}. Baryonic (gas) inflow into the
gravitational potential wells created by these halos builds the
stellar mass and central black holes in the first galaxies
\citep[e.g.][]{Ren77, Fal80, Dal97, Mo98, Mac01, SHD03, Ven06}.
{\color{black}Observational and theoretical work has suggested
that feedback from supernovae and Active Galactic Nuclei (AGN),
powered by the central black holes, may regulate the star
formation in galactic systems \citep[e.g.][]{Kau03, Car05, Nes06,
Scha07, Kav07b, Ala11}} (and perhaps also in systems in their
immediate vicinity, see Shabala et al. 2011) which plausibly
produces the observed correlation between the mass of the central
black hole
and the stellar mass contained in spheroids at present day \citep[e.g.][]{Fer00}.\\
 \indent Galaxy merging {is thought} to be an important driver of the evolution of the visible
Universe \citep[e.g.][]{Nav02}. Mergers {are believed
to drive} strong star formation episodes \citep{Mih96},
{they may contribute to} the growth of black holes
\citep{Spr05} and they're expected to produce morphological
transformations \citep{Too77}. Major mergers between equal mass
progenitors are thought to lead to the formation of early-type
galaxies, largely independent of the original morphologies of the
progenitors \citep[e.g.][]{Whi91, Bar92, Col00}. Repeated minor
mergers appear to be able to produce the same effect
\citep[e.g.][]{Bou07, Naa09}. Observational evidence for the role
of mergers in creating early-type galaxies is suggested by the
presence, in many ellipticals, of morphological disturbances such
as shells, ripples and tidal tails \citep[e.g.][]{b2, Fer09,
Kav10b} and evidence for recent
merger-driven star formation \citep{Kav08,Kav09,Kav10}.\\
\indent Since mergers are expected to play an important role in
the formation of early-type galaxies \citep{Bar96, Ben95}, an
observational study of merger remnants that are likely to end up
as early-type galaxies is desirable. A critical issue in studying
post-mergers (and indeed mergers in general) is that perhaps the
most reliable method for identifying mergers and merger remnants
is by visual inspection of galaxy images. Unfortunately this
technique becomes very impractical for modern large-scale surveys,
such as the \emph{Sloan Digital Sky Survey} (SDSS) which comprises
more than one million galaxies in its spectroscopic sample
\citep{Yor00}. Several automated techniques are able to extract
mergers from survey images, all of which have made significant
contributions to the merger literature but do have some
limitations. For example, merger studies are often based on
samples of `close pairs'\footnote{Close pairs are normally defined
using a projected distance of 30 kpc and a line of sight velocity
differential ($\Delta z \leq$ 500 km s$^{-1}$) \citep{Pat00}} but
involves an inherent (albeit well-motivated!) assumption that the
close-pair system will eventually merge \citep[e.g.][]{Pat00}.
Close-pair techniques can also be biased against minor mergers,
since the smaller merger progenitor is often fainter than the
spectroscopic limit of the survey in question. Similarly, mergers
and merger remnants can be identified via structural parameters
such as `concentration' and `asymmetry' \citep[e.g.][]{Con05}.
While this technique has achieved significant success in probing
the merger population in modern surveys \citep[e.g.][]{Con08}, it
is difficult to define a parameter space uniquely occupied by
mergers and the results typically have to be `calibrated' against
the results of visual inspection \citep{Jog08}.

The Galaxy Zoo \citep[GZ;][]{Lin08} project offers a useful
alternative. GZ has enlisted 300,000+ volunteers from the general
public to morphologically classify, through direct visual
inspection, the entire SDSS spectroscopic sample. This includes
the compilation of a large homogeneous sample of merging systems
in the local Universe \citep{Dar09a, Dar09b}. {The
Darg et al. merger sample was extracted from the SDSS Data Release
6 \citep{Yor00, Ade08}, within the redshift range $0.005<z<0.1$.
At least one of the two galaxies in each merger has $r<17.77$
(which is the SDSS spectroscopic limit). The final sample of
mergers contains 3373 systems, with mass ratios between 1:1 and
1:10. For a more detailed description of the properties of the
sample we refer readers to \cite{Dar09a}.} This merger catalogue
includes a sample of `post-mergers', where the system consists of
a single object, morphologically disturbed as a result of the
recent merger, but in the final stages of relaxation. This study
is based on a subset of these post-mergers in which the
post-merger system has a dominant bulge, making them plausible
progenitors of early-type galaxies.

The plan for this paper is as follows. In Section 2 we discuss the
general properties of the sample. In Section 3 we study the local
environments of our SPMs. In Section 4 we discuss their colours
and emission line activity while in Section 5 we reconstruct the
plausible progenitors of the SPMs. We present our summary in
Section 6.
\begin{figure}
\centering
\includegraphics[width=78mm]{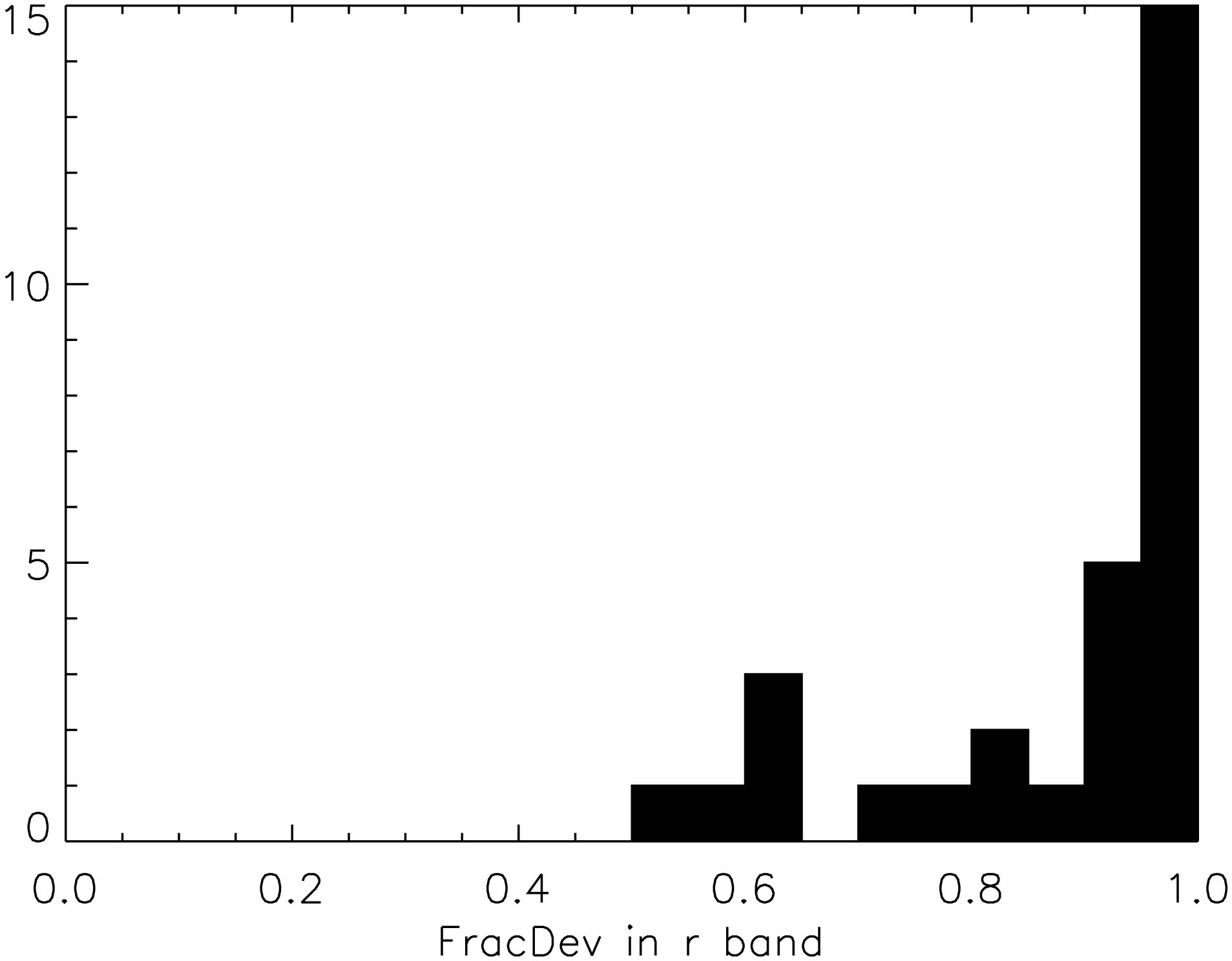}
\caption{\small The \emph{fracdev} parameter in the optical $r$
band for our visually selected sample of spheroidal post-mergers.
All galaxies have \emph{fracdev} higher than 0.5 (most of them
being higher than 0.8) consistent with them being bulge-dominated
systems.}\label{fig:frac}
\end{figure}

\section{The Sample}
We construct a sample of \emph{spheroidal} post-mergers (SPMs)
from a parent sample of 370 post-mergers selected by Darg et al.
As mentioned above, a postmerger is defined as a \emph{single}
object in which the morphological disturbances induced by the
recent merger remain visible. In other words a postmerger
represents the very final stages of the merger process. We
visually re-inspect this postmerger sample to select 30 objects
which are clearly bulge-dominated (Figure \ref{fig:sample}). As a
sanity check of our visual classification, we use the SDSS
parameter \emph{fracdev} in the optical $r$ band (see Table
\ref{tab:Sample}). \emph{Fracdev} indicates the likelihood of the
surface brightness profile to be modelled by a pure de Vaucouleurs
profile (i.e. \emph{fracdev} $\sim$ 1 indicates a pure bulge,
while \emph{fracdev} $\sim$ 0 represents a purely exponential or
disk-like profile). Past work \citep[e.g.][]{Kav07} has
successfully used this parameter as a measure of morphology in
large galaxy samples. In Figure \ref{fig:frac} we plot the
\emph{fracdev} in the $r$ band of our sample. All our SPMs have a
\emph{fracdev} greater than 0.5 (i.e. they are dominated by the
`bulge-like' profile), with most of them higher than 0.8, which is
consistent with the results of our visual classification. Note
that not all postmergers with high values of \emph{fracdev} from
the original sample of 370 were in fact bulge-dominated when
inspected visually, a further example of the utility of employing
visual inspection in conjunction with automatic techniques.\\
\indent {\color{black} Following the estimated completeness of the
Darg et al. merger sample, we expect the completeness of our SPM
sample to be around 80\%.}

\begin{table*}
 \begin{center}
\begin{tabular}{c|c c c c c c c}
\hline
SDSS ID&RA&DEC&FracDev in $r$&$\rho$&Redshift &Log(Mass)\\
\hline
587726102026453183 & 227.770& 4.293 & 0.799 & 0.536& 0.0420& 11.552\\
  587726102027239592&   229.454 &  4.162 & 0.628 &0.609&0.037&11.722\\
  587731521205567501    &132.978   &40.835&1&1.08519& 0.029&11.153\\
  587731681194868806 &  139.212&   45.700&0.948&0.084&0.026&11.073\\
  587732484342415393 &  130.937 &  35.828&0.638&1.942&0.054&11.002\\
  587733431923703839&   253.789 &    26.664&0.940&1.017&0.035&11.163\\
  587734948595236905&   160.265&   11.096&1&1.019&0.053&11.206\\
  587735663161442343&   153.992&   39.243&1&0.547&0.0627&10.865\\
  587735665845403787&   221.437&   51.580&0.894&0.0156&0.030&11.235\\
  587735696984571992&   201.726&   56.889&1&0.207&0.090&11.172\\
  587736585513795652&   244.425&   25.205&0.950&0.043&0.0311&10.294\\
  587736586036969554&   211.414&   40.032&1&0.046&0.084&11.335\\
  587738409785557168&   143.447 &10.811    &0.711&0.00676&0.085&11.758\\
  587738946141552732&   169.385&   37.963 &0.943&0.098&0.096&10.799\\
  587738947747053602&   154.640&   36.224 &0.851& 0.591& 0.054&11.056\\
 587739707951808602&    227.963&   23.151 &1&0&0.052&11.171\\
  587741533323526200&   173.781&   29.891&0.857&0.103&0.046&10.980\\
  587741533859414124&   171.142&   30.095& 1&0.148&0.055&10.722\\
  587741600963690567&   198.656&   26.123&    1& 0.079&0.074&11.251\\
  587741708326469917&   128.289&   15.398 &1&0.219&0.076&10.802\\
  587741709954121847&   168.418&   27.241  &0.658 & 0.453& 0.037&10.263\\
  587742062680015040&   176.183   &23.162    &1&0&0.048 &10.621\\
  587744874792222779&   137.156&   14.122 &1&0.158 &0.0882&11.080\\
  588007005789683827&   196.060&   65.345 &0.598&0.412&0.083&11.087\\
  588013382183944595&   119.951&   27.838 &0.536&0.0442&0.067&10.977\\
  588015508746338312&   31.566&    -0.2914 &1&0.529&0.0426&11.329\\
  588017565490872652&   187.554&   11.770 &0.922& 1.459&0.089&11.012\\
  588017705071214745&   164.196&   12.762  &0.963&0.5035& 0.092&11.421\\
  588017978367934481&   218.326&   34.734 &0.996& 0.288&0.034&11.190\\
  588017978895892603&   193.458&   39.738 &1&0.259&0.092&10.814\\
\end{tabular}
\caption{\small SDSS ID, RA, DEC, $r$-band \emph{fracdev},
environment parameter $\rho$, spectroscopic redshift and stellar
mass for our sample of postmergers. Stellar
masses are estimated using the calibrations of Bell et al. (2003) } \label{tab:Sample}
\end{center}
\end{table*}
\begin{figure}
\centering
\includegraphics[width=86mm]{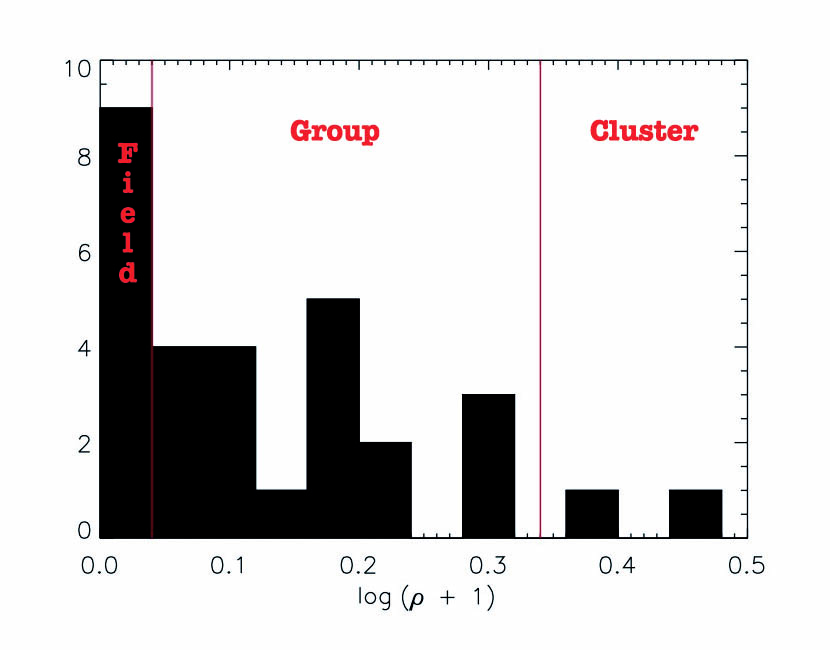}
\caption{\small The values of the environment parameter $(\rho)$
for the sample: two galaxies in our sample inhabit clusters
$(log(\rho +1) > 0.30)$, with the rest split between groups $(0 <
log(\rho +1) < 0.30)$ and the field. See text for details.}
\label{fig:Rho}
\end{figure}
\begin{figure}
\centering
\includegraphics[width=86mm]{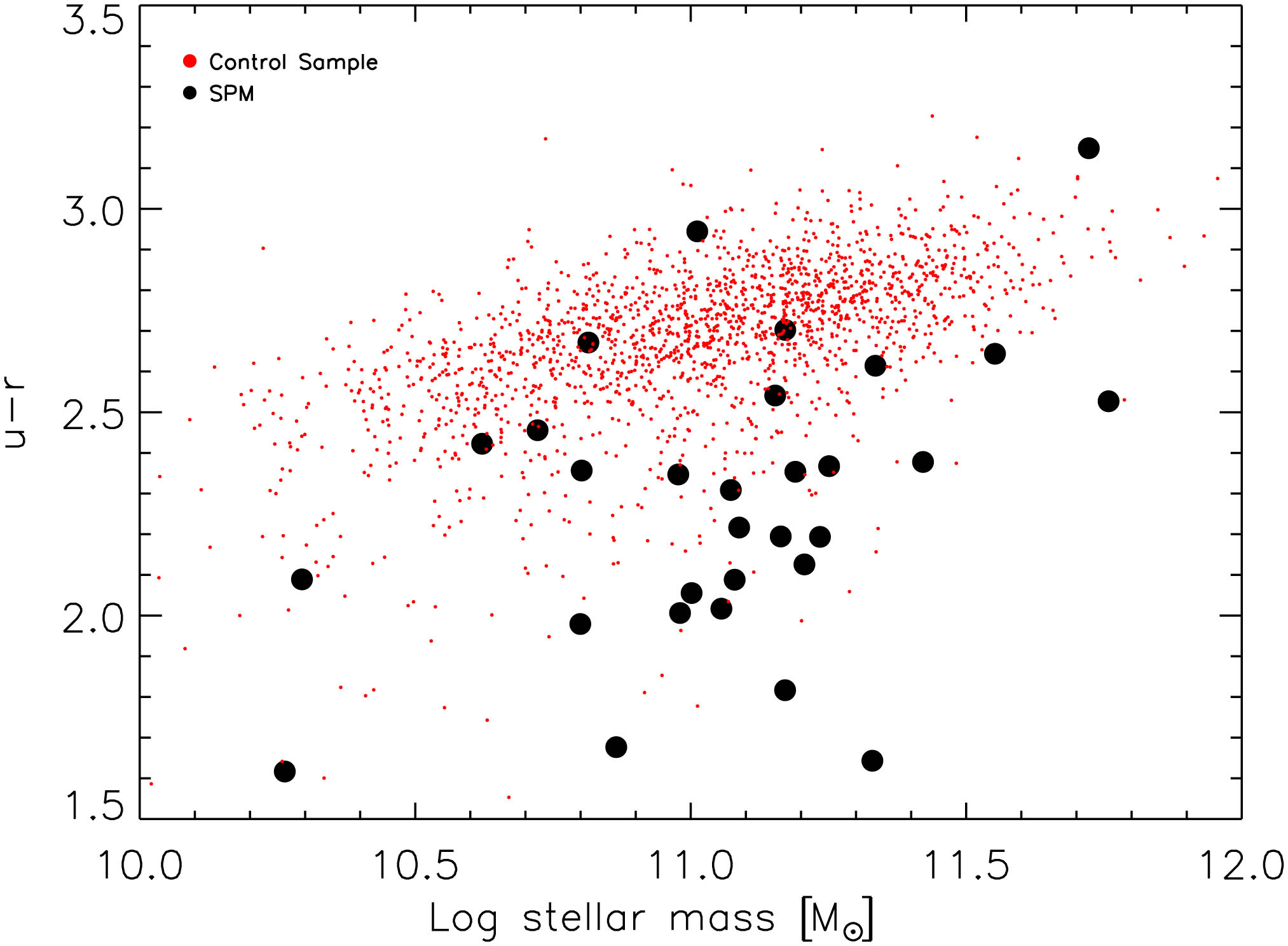}
\caption{\small The $(u-r)$ colour-mass relation of the spheroidal
postmergers (large black circles), compared to a control sample of
early-type galaxies from the SDSS (red dots).}
\label{fig:Colours}
\end{figure}
\section{Local environments}
We begin by exploring the local environments of our SPMs, using
the environment parameter ($\rho_g$) defined by \cite{Sch07}. This
is defined as a weighted sum of all the neighbours within the
ellipse
\begin{equation}
{\big({r_a \over 3 \sigma} \big)} + {\big({r_z \over 3 c_z \sigma} \big)} \le 1
\end{equation}
where $r_a$ is the distance on the sky in Mpc to each surrounding
galaxy, $r_z$ is the distance along the line-of-sight in Mpc to
each surrounding galaxy, and $\sigma$ is the radius. The parameter
$c_z$ scales the value of $\sigma$ along the line of sight to
compensate for the `finger of god' effect (see Schawinski et al.
2007 for more details)\footnote{The finger of god is an effect in
observational cosmology that causes clusters of galaxies to be
elongated in redshift space, with an axis of elongation pointed
towards the observer. It is caused by a Doppler shift associated
with the peculiar velocities of galaxies in a cluster.}. According
to this definition a galaxy with $\rho_g=0$ typically has no
neighbours in a $\sigma$ radius. Values in the range $ 0 \textless
\rho_g \textless 0.1$ are consistent with a field environment.
Galaxies with $0.1 \textless \rho_g \textless 1$ are in a group
environment, while anything larger typically corresponds to
clusters. We find that two galaxies in our SPM sample inhabit
clusters ($\rho_g > 1$), while the rest are split between groups
and the field. This result is expected since the high peculiar
velocities of the galaxies in dense environments such as clusters
make collisions unlikely. Similarly, in very sparse environments
there are not enough galaxies around to produce many merger
events, so a post-merger population spread between intermediate
and low-density environments is reasonable. The environment
parameter values for the sample are shown in Figure \ref{fig:Rho}.
\begin{figure}
\begin{center}
   \includegraphics[width=86mm]{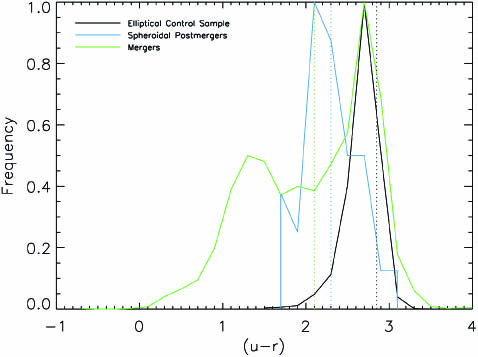}
\caption{\small Histogram of $(u-r)$ colour of the SPMs, ongoing
mergers from the catalogue of Darg et al. and a control sample of
early-type galaxies from the SDSS. The SPMs lie intermediate
between the ongoing mergers and early-type control sample,
indicating that the star formation activity peaks before the
post-merger phase. The dotted lines are the median values for the populations.} \label{fig:HUR}
\end{center}
\end{figure}

\section{Colours and emission line activity}
In Figures \ref{fig:Colours} and \ref{fig:HUR} we compare the
colours of the SPMs to both an early-type control sample and the
\emph{ongoing} mergers from Darg et al. The colours are
$K$-corrected according to the technique devised by \cite{Bla07}
using the IDL routine $\mathtt{KCORRECT}$(version 4.2). Stellar
masses are estimated using the calibrations of \cite{Bell03}. The
early-type control sample is constructed using the GZ early-type
galaxy catalogue restricted to the redshift and magnitude range of
the Darg et al. mergers \citep[see][]{Dar09a}. {The
vast majority (85$\%$) of our SPM sample have bluer colours than
the mean colours of the early-type control population (Figure
\ref{fig:HUR}). These blue populations are likely to contain both young
stars formed in the recent merger as well as remnants of the blue
stellar populations in the original progenitors. However, they
are typically not bluer than the population of ongoing mergers
from Darg et al. (2010), which suggests that the star-formation rate is subsiding in the postmerger phase}. {While there
is some debate on exactly when the star formation activity peaks
during the merger process \citep{Bar00,Lam03,Nic04, Car05,
Sch09b}, our results suggest that it peaks prior to the final
coalescence of the merger progenitors.} We use optical
emission-line ratios \citep[see e.g.][]{BPT81} to explore the
emission-line activity in the SPM sample and compare it to what is
found in the control sample of early-type galaxies and the ongoing
mergers from Darg et al. Emission lines are calculated using the
public GANDALF code\footnote{GANDALF is a simultaneous emission
and absorption lines fitting algorithm, designed to separate the
relative contribution of the stellar continuum and nebular
emission in the spectra of nearby galaxies, while measuring the
gas emission and kinematics. This method has been used to derive
the ionised-gas maps and kinematics of the SAURON sample
\citep{b3}} \citep{b3}.

The majority of the SPMs display Seyfert-like emission (42$\%$)
with the rest being either LINERs (26$\%$), star-forming (16$\%$),
or quiescent (16$\%$) objects. In comparison, the dominant
emission-line type in the ongoing mergers is the star-forming
population, while the dominant type in the control early-types are
quiescent objects. Together with the higher fraction of LINERs in
the SPMs (which are likely to be post-starburst galaxies, see e.g.
\cite{b7,Sar10}) {and since it has not been possible
to identify blue Compton-thick AGN in the local universe
\citep{Sch09b} our results suggest that these objects have gone
through a gradual transition from being dominated by star
formation in the merger phase to AGN activity in the post-merger
phase, followed by quiescence when the objects have transitioned
to being relaxed spheroids}. While the small number of SPMs makes
a robust conclusion difficult, our results suggest that, not only
is there a delay between the onset of star formation and AGN
activity, in agreement with several studies in the literature
\citep{Scha07, Wild10,Dar09b}, \emph{but that the peak of AGN
activity may coincide with the post-merger phase of the merger
process.} Note that, in this case, the delay between the onset of
star formation and AGN activity could be expected to be around the
coalescence timescale of the merger, which is around 0.5-1 Gyr for
a major merger \citep{Spr05, Lot08}. This appears consistent with
the time delays derived from spectral fitting in recent work
\citep{Scha07, Wild10, Dar09b}.

\section{Reconstructing the progenitors}

\begin{figure*}
\centering
\includegraphics[width=150mm]{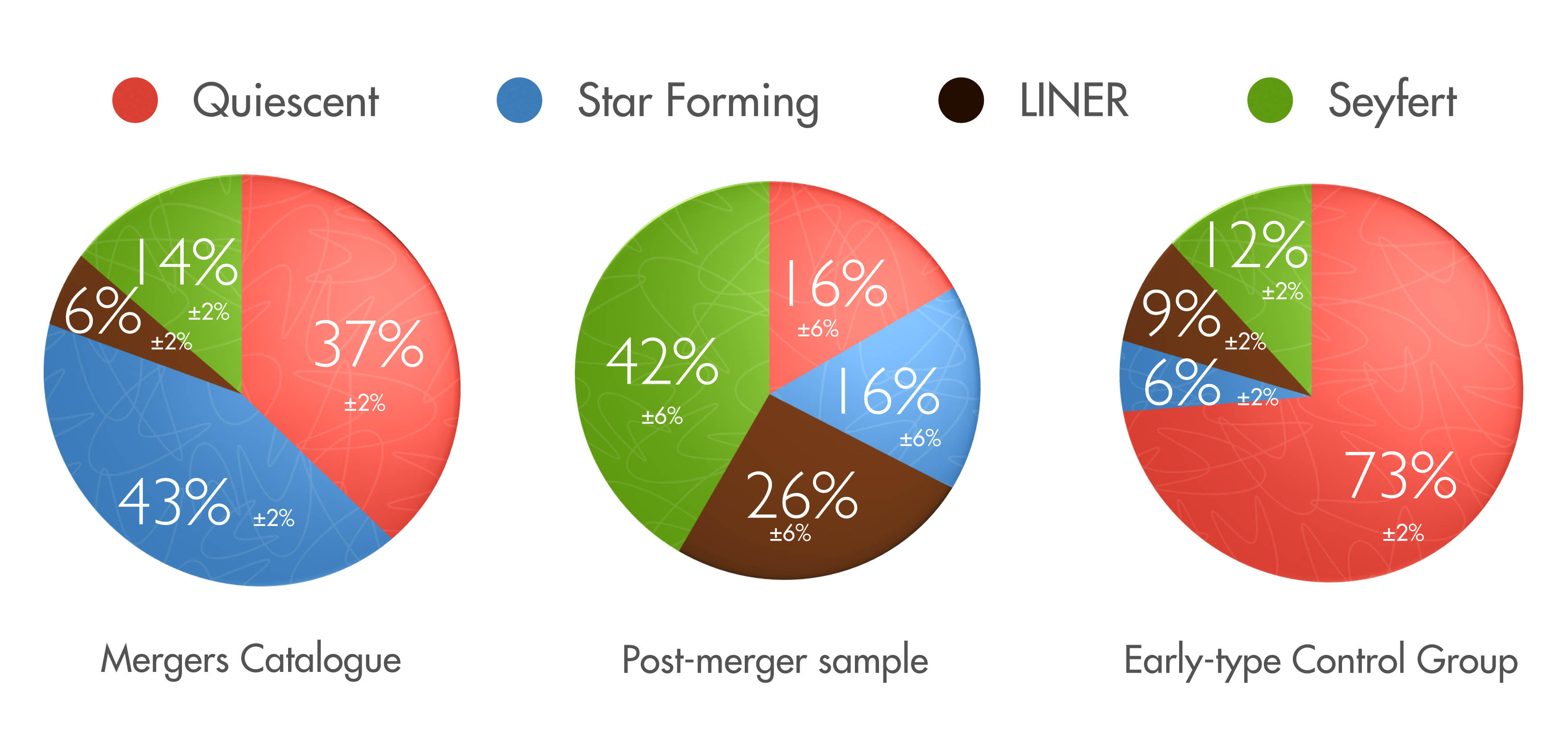}
\caption{Emission-line analysis of the SPMs compared to the
ongoing mergers from Darg et al., and a control sample of
early-type galaxies from the SDSS. The dominant emission-line type
in the ongoing mergers are star-forming galaxies, while the
dominant type in the SPMs and early-type controls are AGN and
quiescent galaxies respectively. The AGN phase appears to dominate
the postmerger phase in the morphological sequence.}
\label{fig:Comp}
\end{figure*}

\begin{figure*}
\centering
\includegraphics[width=130mm]{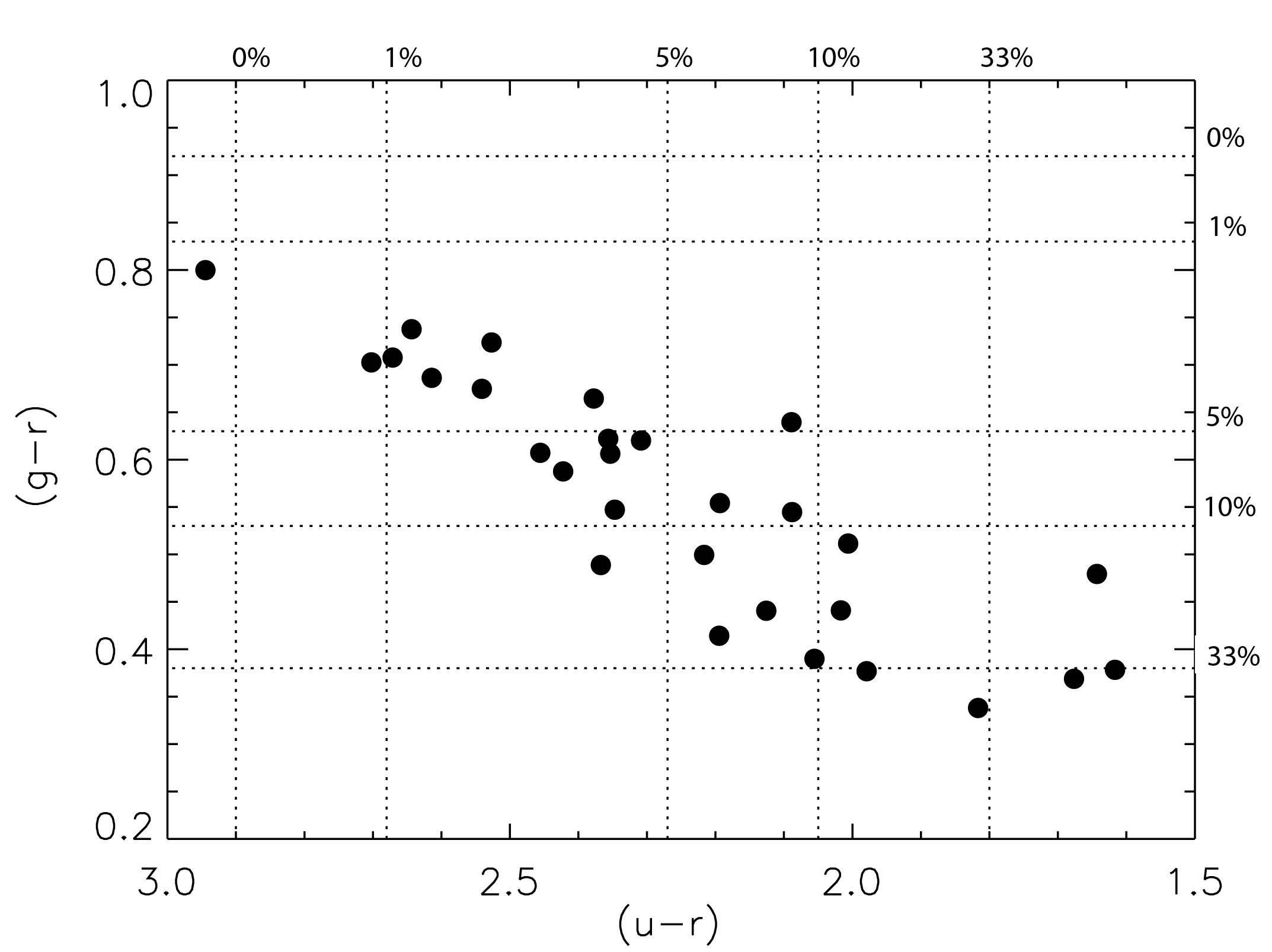}
\caption{\small Comparison of the colours of the spheroidal
postmergers to a simple model, designed to approximate the stellar
content in these galaxies. The model is parametrised by two
instantaneous starbursts, the first fixed at an age of 10 Gyr
(since the bulk of the stars in early-type galaxies form at high
redshift). The second burst is assumed to take place 0.5 Gyr in
the past (the typical coalescence timescale of major mergers). The
mass fraction contributed by this young population is indicated by
the horizontal and vertical lines. Note that we assume solar
metallicity and no dust in the models.}
\label{fig:gr}
\end{figure*}

In this section we explore the plausible progenitors of the SPM
sample. Under the reasonable assumption that the sample of ongoing
mergers in Darg et al. are the progenitors of the SPMs, we first
search for merging systems which have a summed mass within 0.3 dex
of the mass of the SPM in question. We use 0.3 dex as the
tolerance because it is the typical mass error. We assume that
major mergers of all morphological types (elliptical-elliptical,
elliptical-spiral and spiral-spiral) produce spheroids
\citep[e.g.][]{Kho06} and that, from simple dynamical
considerations, all minor mergers whose major partner is an
elliptical will create a spheroid. Note that the merger catalogue
of Darg et al. is not expected to be biased against minor mergers
with mass ratios between 1:4 and 1:10. The median mass ratios for
the SPMs, under these assumptions, are between 1:1.5 to 1:3, with
a tail to lower values, suggesting that these systems are
typically remnants of major mergers.

To explore the morphologies of the progenitors further, we make a
simple comparison of the SPM colours to the expected colour of a
passively evolving old population, typical of early-type galaxies
at the present day \citep[e.g.][]{Bow92, Ben97, Spr05, Del06,
Kor09}. The colours of the SPMs suggest that, in most of these
galaxies, a significant minority of the stellar population is
likely to have formed in the recent merger-driven burst of star
formation.

{To explore the star formation histories of the SPMs
further, we study simple models in which a young population,
formed in an instantaneous starburst, is superimposed on an old
population that is also formed in an instantaneous event. We
assume that the old population is formed at $z=3$ and has solar
metallicity, since this is typical of the old, metal-rich stellar
populations that dominate nearby early-type galaxies
\citep[e.g.][]{Tra00}. The free parameters are the
age, mass fraction and metallicity of the recent (young) starburst
and the average dust content (E$_{(B-V)}$) of the system. We
explore ages for the recent starburst between 0.2 and 0.6 Gyrs,
which bracket the coalescence timescales for major mergers in the
literature \citep{Spr05, Lot08}. We explore metallicities between
0.75 and 2.5 Z$_{\odot}$ which is the scatter in the
mass-metallicity relation of galaxies in our mass range in the
local Universe \cite{Tre04}. We assume a median E$_{(B-V)}$ of
0.05, derived by recent UV-optical studies of nearby early-type
galaxies \citep{Kav07b, Scha07}.

Early-type galaxies are largely devoid of gas. Gas fractions in
early-types within the mass range considered in this study are
typically (much) less than 5\% \citep{Kan04,You02}. Mergers that have two early-type progenitors are therefore very
unlikely to produce more than 5\% in mass fraction of young stars.
By comparing the $(u-r)$ and $(g-r)$ colours of the simple models
with the observed colours of the SPMs, we estimate how many of our
postmergers may be the product of mergers between two early-types
and therefore how many are likely to require at least one
late-type progenitor. Considering the range of
parameters described above we find that $\sim$55$\%$ are likely to
have formed in a merger involving a gas-rich i.e. late-type
galaxy. Figure \ref{fig:gr} demonstrates this analysis for a
specific model in the young starburst has an age of 0.5 Gyr, solar
metallicity and an E$_{(B-V)}$ of 0.05.}

\section{Summary}
We have studied a sample of 30 bulge-dominated or spheroidal
post-mergers (SPMs) in the local Universe which are, by virtue of
their morphology, plausible progenitors of early-type galaxies.
These galaxies are a subset of Darg et al. (2010a)) who have
produced a large, homogeneous catalogue of mergers, through direct
visual inspection of the entire SDSS spectroscopic sample using
the Galaxy Zoo project.

The vast majority of the SPMs inhabit low-density environments
(groups and the field), consistent with the expectation that the
high peculiar velocities in high-density environments make
conditions difficult for galaxy merging. Our SPM sample is
generally bluer that a control sample of early-type galaxies but
redder, on average, than the merging population, indicating that
the peak of star formation activity takes place during the merger
phase. 84$\%$ of the SPMs exhibit emission-line activity. 42$\%$
show Seyfert-like emission, 26$\%$ are LINERs and 16$\%$ are
classified as star-forming. In contrast, the control sample of
early-type galaxies is dominated by quiescent objects, while the
mergers are dominated by star-forming galaxies. The rise in the
AGN fraction in the post-merger phase (compared to the mergers)
suggests that the AGN phase probably becomes dominant only in the
very final stages the merging process. Comparison of the SPMs to
the ongoing mergers in the Darg et al. sample indicates that they
are likely to be the remnants of major mergers.\\ \indent  Since major
mergers coalesce on timescales around 0.5 Gyr, we have compared
the colours of the SPMs to models in which a young stellar
population with an age of 0.5 Gyr is superimposed on an old
population that forms at $z=3$ (since the bulk of the stars in
early-type galaxies are known to be old).We have found that,
under these assumptions, the vast majority of the SPMs are likely
to have formed more than 5\% of their stellar mass in the recent
merger driven burst. {Since early-type galaxies
themselves are rather gas-poor objects, our results indicate that
$\sim$55\% of the SPMs are products of major mergers in which at
least one of the progenitors is a late-type galaxy.}

\section*{Acknowledgments}
\indent SK acknowledges support from the Royal Commission for the
Exhibition of 1851, Imperial College and Worcester College,
Oxford. We thank David Clements and Yvonne Unruh for many
constructive comments.

This publication has been made possible by the participation of
more than 300,000 volunteers in the Galaxy Zoo project. Their
contributions are individually acknowledged at
http://www.galaxyzoo.org/Volunteers.aspx.

Funding for the SDSS and SDSS-II has been provided by the Alfred
P. Sloan Foundation, the Participating Institutions, the National
Science Foundation, the U.S. Department of Energy, the National
Aeronautics and Space Administration, the Japanese Monbukagakusho,
the Max Planck Society, and the Higher Education Funding Council
for England. The SDSS Web Site is http://www.sdss.org/. The SDSS
is managed by the Astrophysical Research Consortium for the
Participating Institutions. The Participating Institutions are the
American Museum of Natural History, Astrophysical Institute
Potsdam, University of Basel, University of Cambridge, Case
Western Reserve University, University of Chicago, Drexel
University, Fermilab, the Institute for Advanced Study, the Japan
Participation Group, Johns Hopkins University, the Joint Institute
for Nuclear Astrophysics, the Kavli Institute for Particle
Astrophysics and Cosmology, the Korean Scientist Group, the
Chinese Academy of Sciences (LAMOST), Los Alamos National
Laboratory, the Max-Planck-Institute for Astronomy (MPIA), the
Max-Planck-Institute for Astrophysics (MPA), New Mexico State
University, Ohio State University, University of Pittsburgh,
University of Portsmouth, Princeton University, the United States
Naval Observatory, and the University of Washington.

GALEX (Galaxy Evolution Explorer) is a NASA Small Explorer,
launched in April 2003, developed in cooperation with the Centre
National d'Etudes Spatiales of France and
the Korean Ministry of Science and Technology.\\

\nocite{Sha11}

\label{lastpage}

\end{document}